\documentclass[a4paper,
               keeplastbox,   
               ]{jacow}
\usepackage{pdfpages,multirow,ragged2e} %
\makeatletter%
	\ifboolexpr{bool{xetex}}
	 {\renewcommand{\Gin@extensions}{.pdf,%
	                    .png,.jpg,.bmp,.pict,.tif,.psd,.mac,.sga,.tga,.gif,%
	                    .eps,.ps,%
	                    }}{}
\makeatother

\ifboolexpr{bool{xetex} or bool{luatex}} 
 {}                                      
 {\usepackage[utf8]{inputenc}}           

\usepackage[USenglish]{babel}

\ifboolexpr{bool{jacowbiblatex}}%
 {%
  \addbibresource{jacow-test.bib}
  \addbibresource{biblatex-examples.bib}
 }{}
\begin{document}

\title{Evaluation of flux expulsion and flux trapping sensitivity of SRF cavities fabricated from cold work Nb sheet with successive heat treatment\thanks{ Work supported by the U.S. Department of Energy, Office of Science, Office of Nuclear Physics under contract DE-AC05-06OR23177 and U.S.Department of Energy, Office of Science, Office of High Energy Physics under Awards No. DE-SC 0009960. }}

\author{B. D. Khanal\textsuperscript{1} and P. Dhakal\textsuperscript{2}\thanks{dhakal@jlab.org} \\
				\textsuperscript{1}Center for Accelerator Science, Department of Physics,\\ Old Dominion University, Norfolk, VA 23529, USA\\
                \textsuperscript{2}Thomas Jefferson National Accelerator Facility, Newport News, VA 23606, USA \\ 
		}
	
\maketitle

\begin{abstract}

The main source of RF losses leading to lower quality factor of superconducting radio-frequency cavities is due to the residual magnetic flux trapped during cool-down. The loss due to flux trapping is more pronounced for cavities subjected to impurities doping. The flux trapping and its sensitivity to rf losses are related to several intrinsic and extrinsic phenomena. To elucidate the effect of re-crystallization by high temperature heat treatment on the flux trapping sensitivity, we have fabricated two 1.3 GHz single cell cavities from cold-worked Nb sheets and compared with cavities made from standard fine-grain Nb. Flux expulsion ratio and flux trapping sensitivity were measured after successive high temperature heat treatments. The cavity made from cold worked Nb showed better flux expulsion after 800 $^\circ$C/3h heat treatment and similar behavior when heat treated with additional 900 $^\circ$C/3h and 1000 $^\circ$C/3h. In this contribution, we present the summary of flux expulsion, trapping sensitivity, and RF results. 
\end{abstract}

\section{Introduction}
Niobium has been the material of choice for superconducting radio-frequency (SRF) cavities not only because of low power loss at the inner surface of the cavities’ inner wall but also its high ductility which makes easier to fabricate the complex structures \cite{hasan}.The niobium is elemental superconductor with highest critical temperature, $T_c \sim $ 9.25 K and highest critical field, $H_c \sim $ 200 mT. The performance is measured in terms of the quality factor ($Q_0$) which defined as the ratio of stored energy inside the cavities to the power dissipation on the inner wall of the cavities per radio frequency (RF) cycle as a function of accelerating gradient ($E_{acc}$).
The ambient magnetic flux trapping during the cooldown is one of the prominent factors causing the degradation of quality factor in cavities. The trapped flux in the form of vortices oscillates in the presence of RF field and dissipate energy. The field dependence of RF losses due to trapped vortices is much stronger than the ohmic-type loss \cite{gigialex08,khanalsrf23}. The ambient flux trapping and the flux trapping sensitivity to rf losses are related to several extrinsic and intrinsic phenomena. The primary host sites of flux trapping are the materials defects, dislocations, impurities, normal conducting precipitates. For instance, the RF loss due to flux trapping can be minimized by maximizing the flux expulsion when the cavity transition to the superconducting state from normal conducting during the cavity cool-down by creating a large thermal gradient across the cavity surface.  Intrinsically, we can minimize the flux trapping by minimizing the defects, dislocations, impurities with different temperature heat treatments followed by chemical and mechanical polishing and by high pressure rinse with de-ionized water. It has been demonstrated that several different pinning mechanism plays a role to the rf losses due to vortices \cite{dhakal20}. Studies showed that doped cavities are more vulnerable to the vortex dissipation loss due to the presence of the dopant on cavities rf surface \cite{sam, dan1, martina1}. The flux expulsion can be maximized by increasing the annealing temperature \cite{sam1}. The increase in annealing temperature minimizes the pinning centers by removing the clusters of dislocations and impurities. In addition, the metallurgical state with larger grain size is expected as the annealing temperature is increased. Fine-grain recrystallized microstructure with an average grain size of 10–50 $\mu$m leads to flux trapping even with a lack of dislocation structures in grain interiors \cite{shreyas}. Thus, it is important to consider the crystallize structure of the niobium before the fabrications and during the cavity processing \cite{antoine}.  
In this contribution, we have fabricated two single cell cavity, one from cavity grade SRF Nb with grain size specified to ASTM 4-6 and other from the cold worked sheet with no specified grain size. The cavity were processed together for chemical polishing with electropolishing and successive annealing at 800, 900 and 1000 $^\circ$C/3h heat treatment. The flux expulsion ratio, flux trapping sensitivity and $Q_0(B_p)$ at 2.0 K were measured. 

\section{Fabrication and Surface Preparation}
The SRF grade and cold-worked Nb sheets with residual resistivity ratio $\ge$ 300 were purchased from Ningxia OTIC, China and 1.3 GHz TESLA shaped cavities were fabricated at Zanon Research \& Innovation Srl, Italy using standard practice of deep drawing to half cells, trimming, machining of the iris and equator of the half-cells, electron beam welding of the beam tubes (made from low purity niobium). The cavity labeled TE1-05 was made from SRF grade Nb and TE1-06 was made from cold-worked sheet. The cavities' inner surface of $\sim$ 150 $\mu m$ was removed in horizontal electropolishing setup using a mixture of electronic grade $HF:H_2SO_4$ = 1:9 at a constant voltage of $\sim 14$ V, temperature of 15 - 20 $^\circ$C and a speed of 1 rpm at Jefferson Lab.

Additional $\sim $25 $\mu$m inner surface was removed as a final surface preparation using electropolishing after each successive heat treatments at 800 $^\circ$C, 900 $^\circ$C and 1000 $^\circ$C for 3 hours. 

\section{Experimental setup}
\begin{figure}[htb]
\centering
\includegraphics*[width=.9\columnwidth]{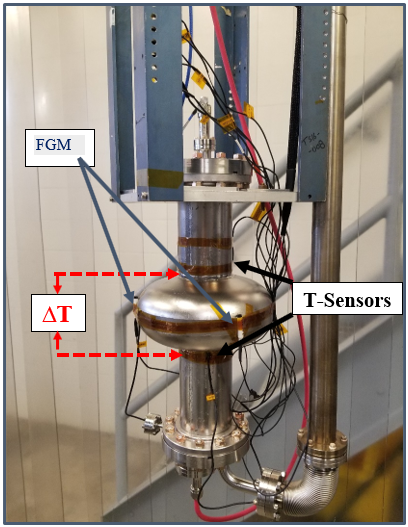}
\caption{Experimental set up with FGMs and temperature sensors.}
\label{fig:setup}
\end{figure}

\begin{figure}[htb]
\centering
\includegraphics*[width=.95\columnwidth]{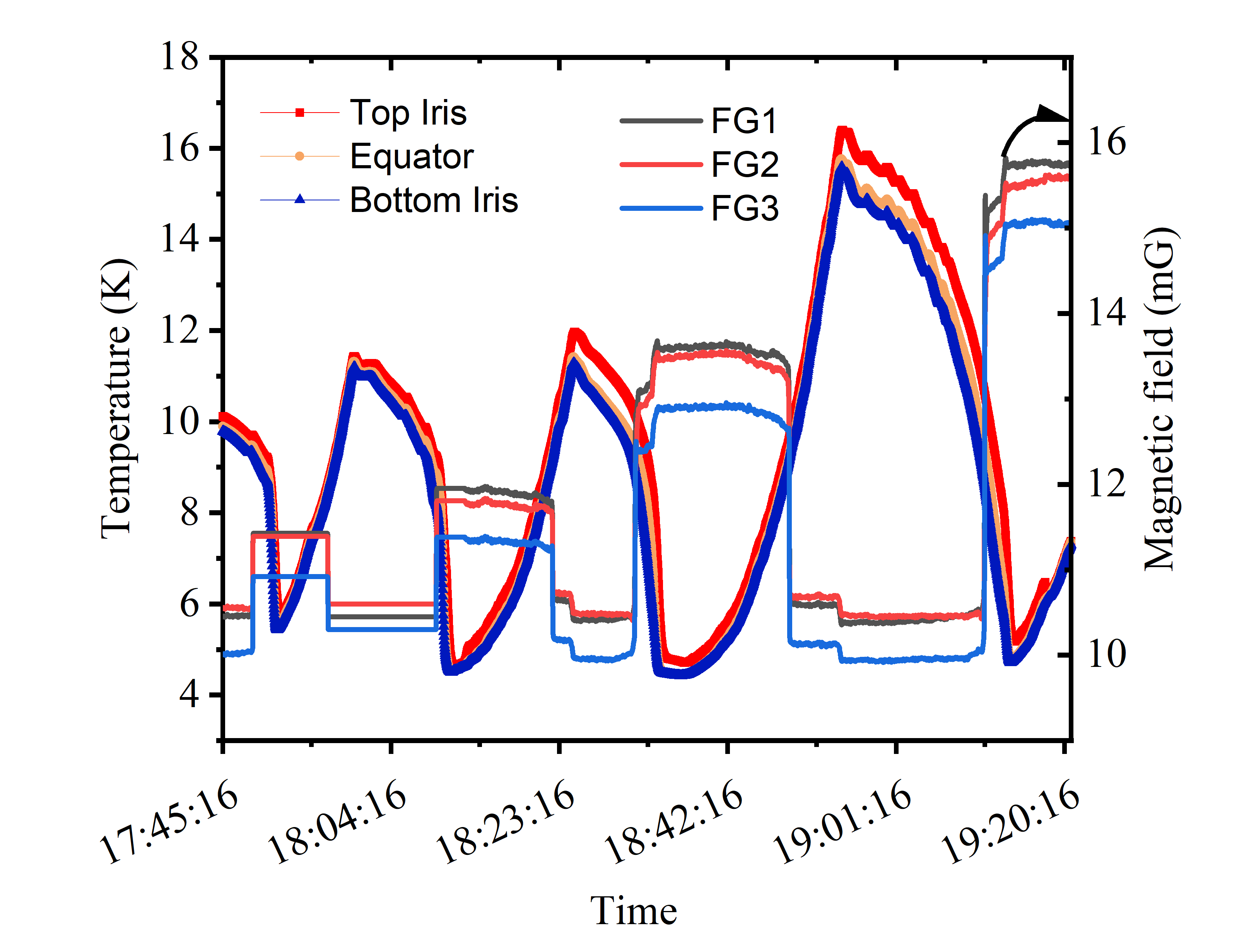}
\caption{The temperature and magnetic field during the cavity cooldown. Depending on the temperature difference between iris-iris, the jump in magnetic field is changed.}
\label{fig:cooldown}
\end{figure}

After the clean room assembly and leak check done, the cavity was transferred to the vertical staging area. Three flux-gate magnetometers (FGM) were fastened at the equator 120$^\circ$ apart each and parallel to the axis of the cavity and six temperature sensors: two sensors at top iris, two sensors at equator, and two sensors at bottom iris were attached. All temperature sensors were attached 180 degree apart as shown in Fig. \ref{fig:setup}. 
The cavity was moved to the Dewar and the temperature and magnetic field was monitored with Cryocon 18i and flux gate magnetometer, respectively during the thermal cycle (cool down and warm up above 10 K). The thermal cycles were done at constant residual magnetic field of $\sim$ 10 mG with the help of compensation coil as shown in Fig. \ref{fig:cooldown}. Several thermal cycles were performed in order to measure the flux expulsion ratio.

After the flux expulsion measurement, the cavity is filled with liquid helium with residual magnetic field 0 and 20 mG in Dewar with cooldown condition of full flux trapping and minimizing flux expulsion by changing the thermal gradient across the cavity surface (iris-iris). The RF measurements consists of $Q_0(T)$ at low peak RF field $B_p\sim 15 mT$ at $Q_0(B_p)$ at 2.0 K.
\section{Results and Discussions}
\subsection{Flux expulsion ratio}
The flux expulsion ratio is defined as the ratio of magnetic field measured in superconducting state ($B_{sc}$) to the magnetic field measured in normal conducting state $B_{n}$ as;
\begin{center}  
Flux expulsion ratio =$\frac{B_{sc}}{B_n}$
\end{center}
The flux expulsion ratio of 1 corresponds to the full flux trapping and ratio of $\sim 1.7$ measured at outer surface of cavity's equator corresponds to the full flux expulsion. The ratio 1.7 calculated by COMSOL simulation assuming a complete Meissner state \cite{dhakal20}.

\begin{figure}[htb]
\centering
\includegraphics*[width=.95\columnwidth]{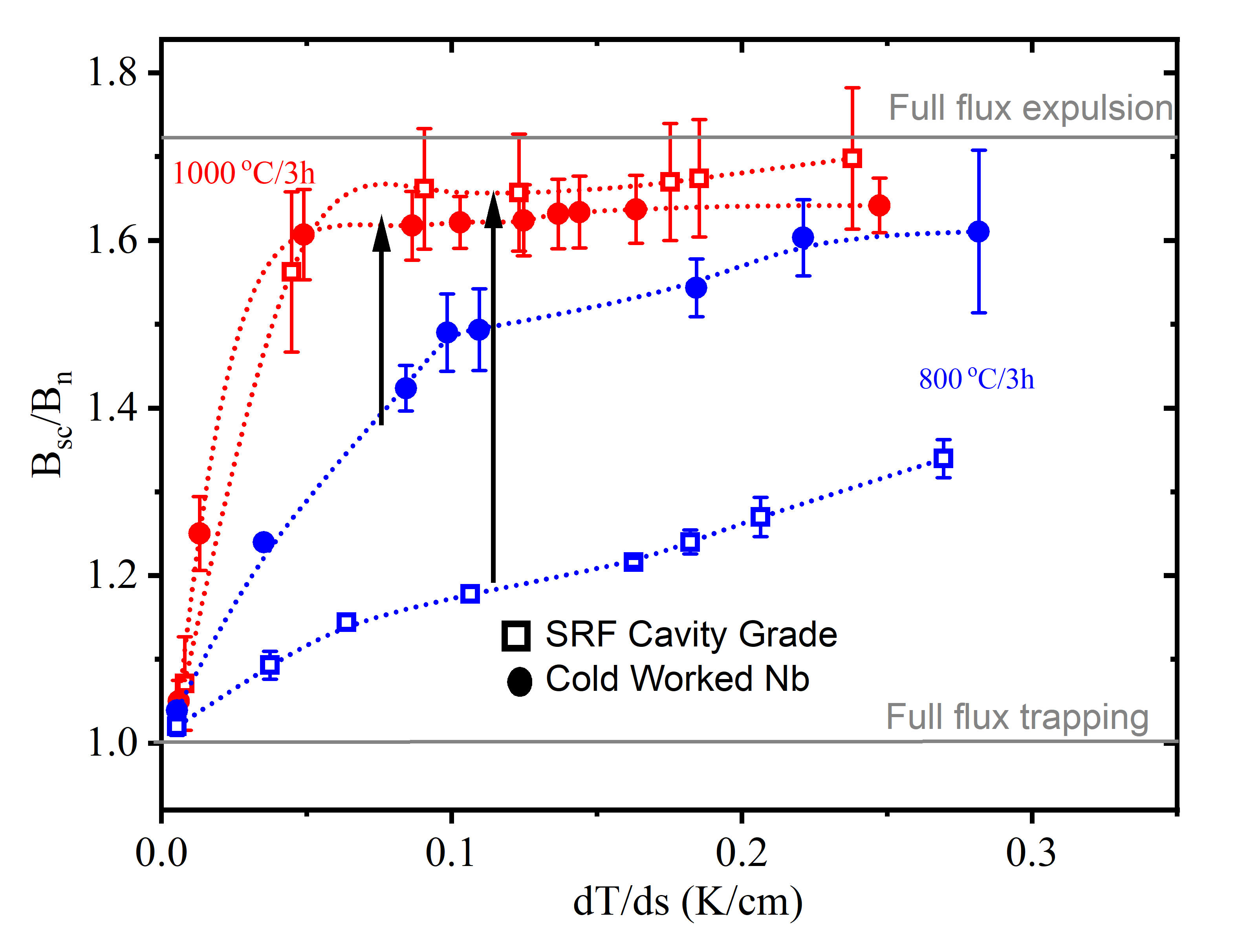}
\caption{Flux expulsion ratio as a function of temperature gradient across the cavity surface (iris-iris) for cavity under study after 800 $^\circ$C/3h and additional 900 $^\circ$C/3h and 1000 $^\circ$C/3h annealing.}
\label{fig:fluxexpulsion}
\end{figure}

Figure \ref{fig:fluxexpulsion} shows the flux expulsion ratio ($B_{sc}/B_n$) as a function of the temperature gradient (dT/ds) along the cavity surface. The cavity made from cold work Nb showed better flux expulsion compared to cavity made from SRF grade Nb after 800 $^\circ$C/3h heat treatment. After additional annealing with 900 $^\circ$C/3h (not shown here) and 1000 $^\circ$C/3h, the flux expulsion ratio for both cavities is similar. The flux expulsion ratio quickly approaches to full expulsion limit for dT/ds > 0.05, which corresponds the temperature difference between the cavity irises to be  $\sim$ 1 K. 

\subsection{RF Results}
\begin{figure}[h]
\centering
\includegraphics*[width=\columnwidth]{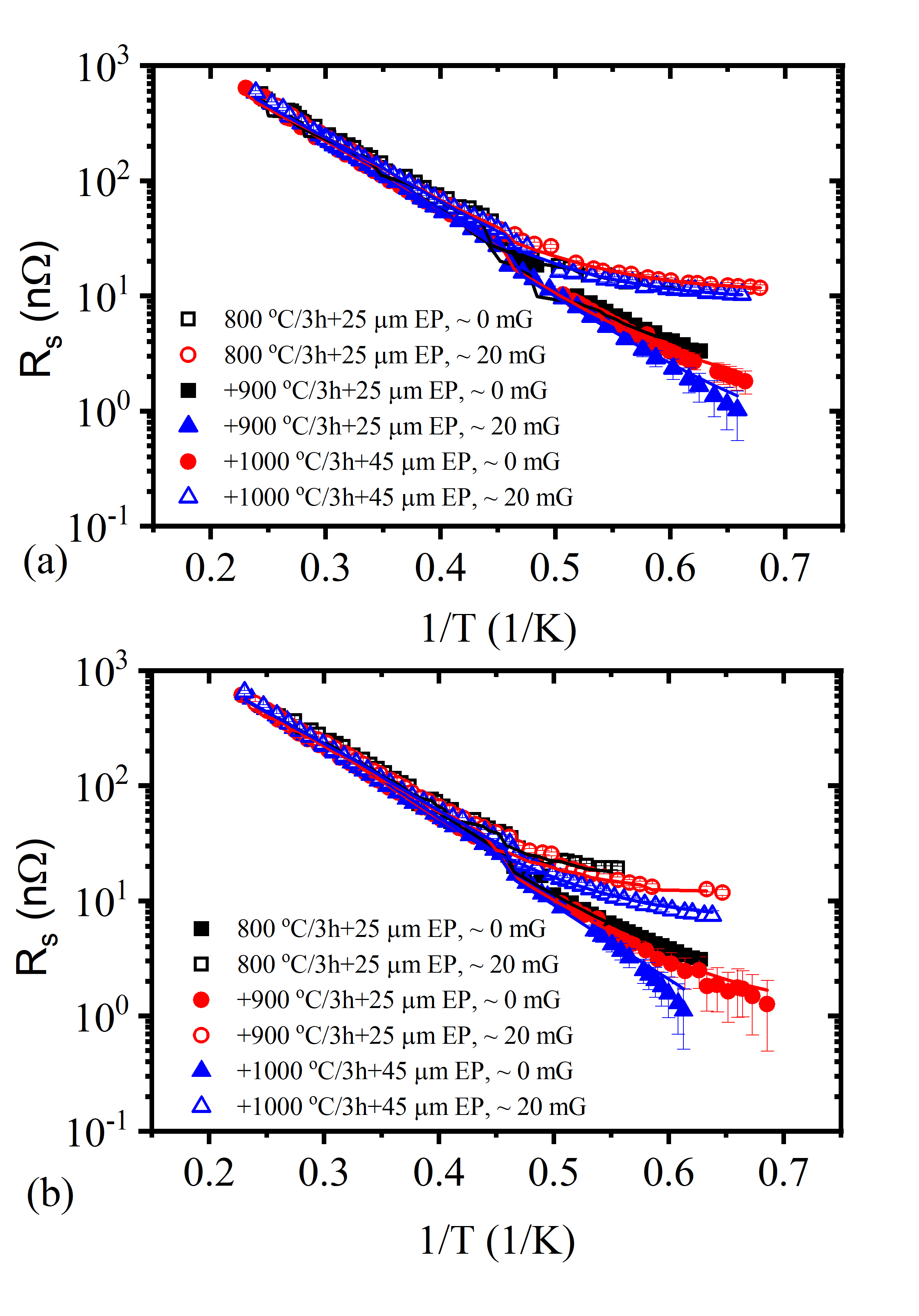}
\caption{$R_s$ vs. 1/T for cavity (a) TE1-05 and (b) TE1-06 after different annealing and cooldown conditions. The solid lines are fits of the $R_s(T)$ using the model described in Ref. \cite{gigi14}.}
\label{fig:rsT1}
\end{figure}

Figure \ref{fig:rsT1} shows the $R_s$ as a function of (1/T) for cavity TE1-05 and TE1-06 for two different cooldown condition with different residual magnetic field in Dewar. The cavity was cooldown through $T_c$ with temperature gradient dT/ds > 0.2 when the residual field in Dewar is $\sim$ 0 mG, where as the cooldown through $T_C$ is kept dT/ds $\sim$ 0 when the residual magnetic field is $\sim$ 20 mG.  The $R_s(T)$ data were fitted using the model developed in Ref. \cite{gigi14} to extract the temperature dependent $R_{BCS}$ and temperature independent residual resistance $R_i$. 
The flux trapping sensitivity is calculated using,
\begin{equation}
    S = \frac{R_{i,B_2}-R_{i,B_1}} {B_2-B_1}.
\end{equation}
where $B_1 \approx $ 0 mG and $B_2 \approx $ 20 mG.

\begin{figure}[h]
\centering
\includegraphics*[width=\columnwidth]{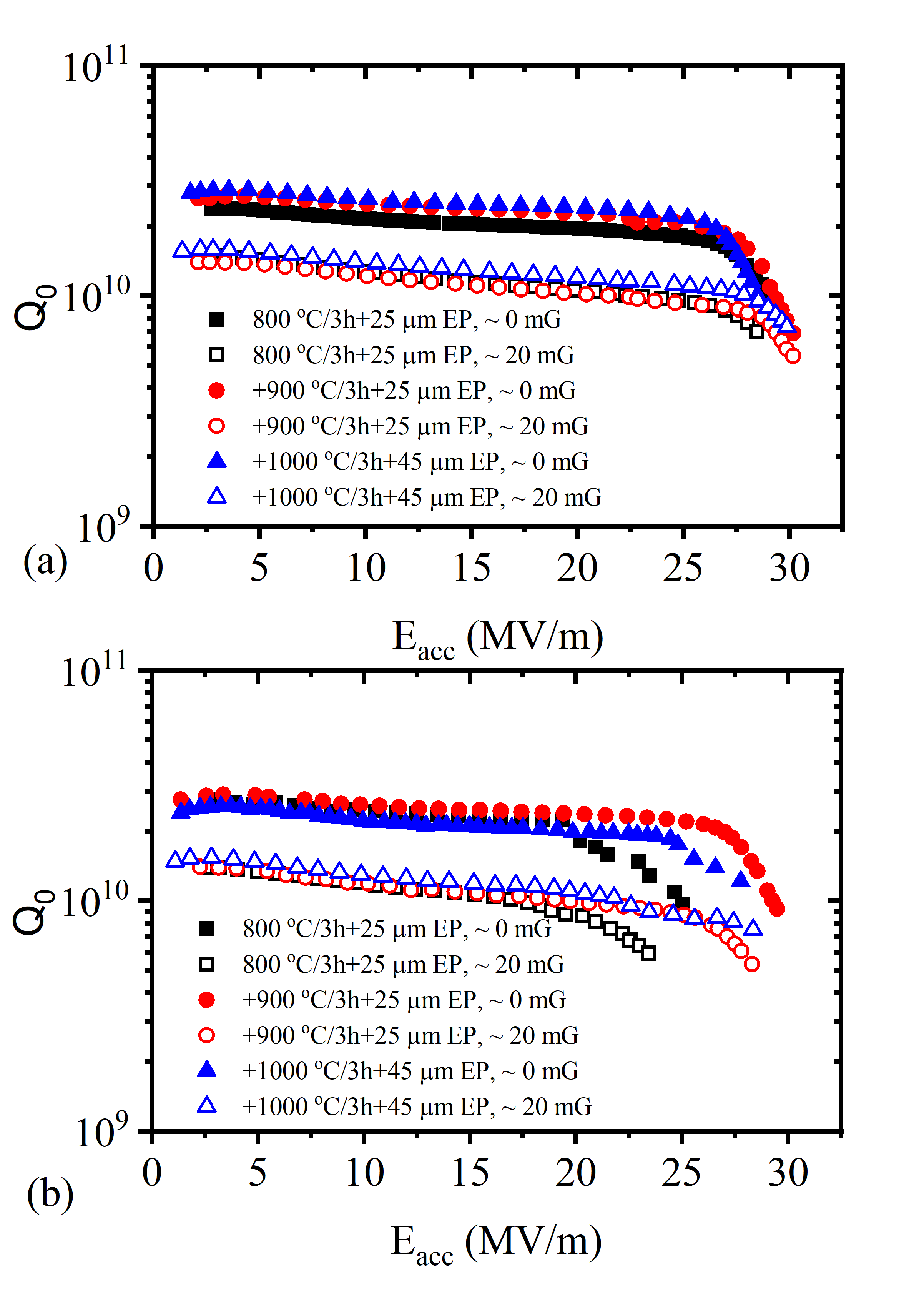}
\caption{$Q_0$ vs. $E_{acc}$ at 2.0 K for cavity (a) TE1-05 and (b) TE1-06 after different annealing and cooldown conditions. All RF tests were limited by quench. The error in $Q_0$ and $E_{acc}$ is < 10 \% and < 5 \% respectively.}
\label{fig:QE}
\end{figure}

Figure \ref{fig:QE} shows the $Q_0$ as a function of $E_{acc}$ at 2.0 K for both cavities with different annealing and cooldown conditions. As expected, $Q_0(E_{acc})$ shows the high field Q-slope starting $\sim$ 26 MV/m. All test RF were limited by quench. Both cavities showed similar performance. The cavity TE1-06 showed some multipacting $\sim$ 20-25 MV/m after 800 and 900 $^\circ$ heat treatments. However, no multipacting were observed after 1000 $^\circ$ heat treatments. It is to be noted that the final surface preparation for all RF tests were $\sim 25~ \mu m$ electropolishing. 

\begin{figure}[htb]
\centering
\includegraphics*[width=\columnwidth]{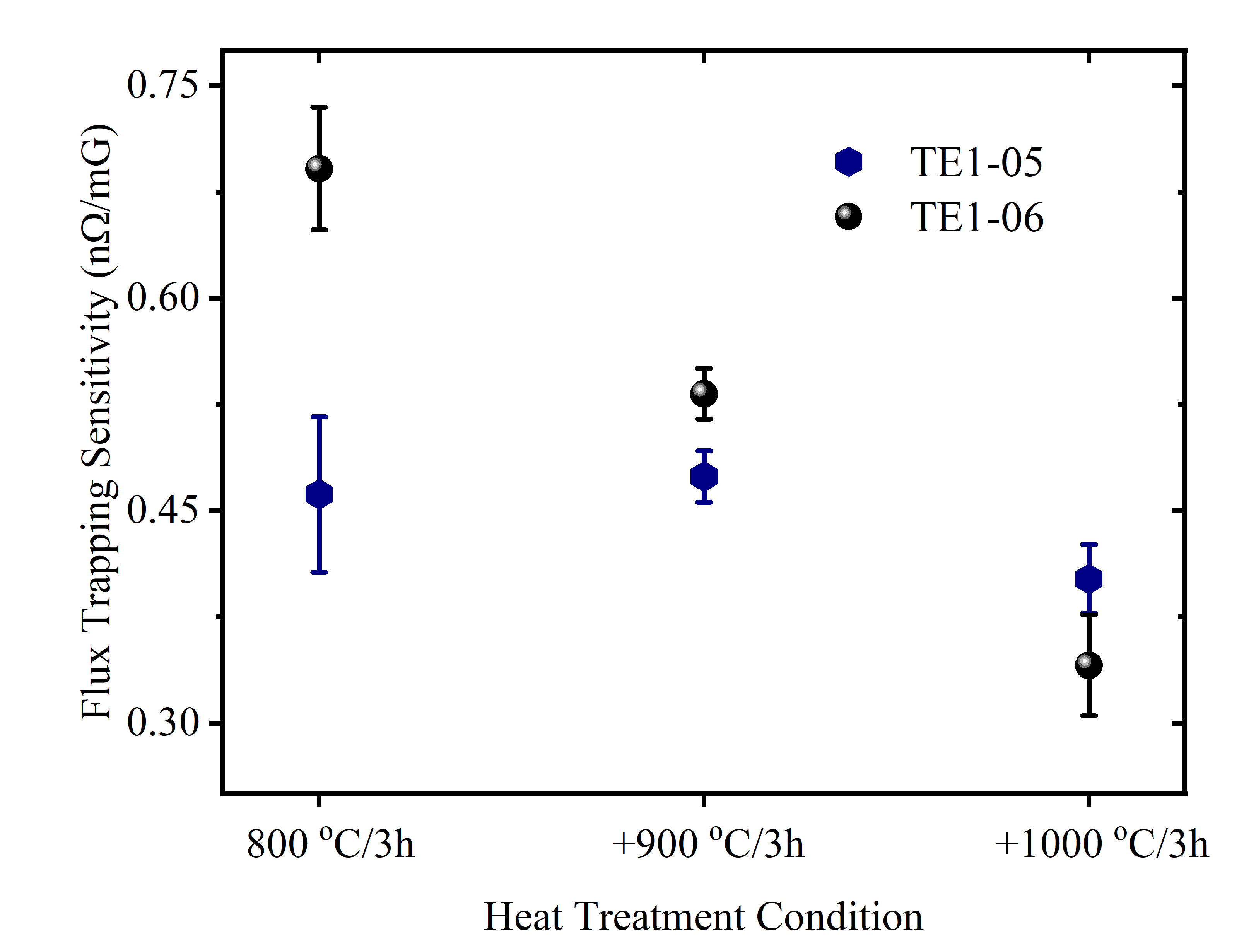}
\caption{Flux trapping sensitivity as a function of annealing conditions. The final surface preparation is electropolishing for all tests.}
\label{fig:sensitivity}
\end{figure}

Figure \ref{fig:sensitivity} shows the flux trapping sensitivity as a function of cumulative annealing temperature. The flux trapping sensitivity decreases with the increase in annealing temperature. The largest change in sensitivity was observed for cavity TE1-06. The flux trapping sensitivity for TE1-06 is higher than TE1-05 after initial 800 $\circ$C/3h heat treatment. After additional 900 and 1000 $\circ$C/3h heat treatment, the sensitivity is very similar.
 \section{Discussion}
As shown in Fig. \ref{fig:fluxexpulsion}, the flux expulsion for cavity made from cold work Nb showed a better expulsion when the cavity was heat treated at 800 $^\circ$C/3h compared to SRF grade Nb. After additional 900 $^\circ$C/3h and 1000 $^\circ$C/3h  heat treatment, the initial microstructure state of the cavity sheet no longer affects the flux expulsion performance \cite{Khanalnapac22}. On the other hand, the flux trapping sensitivity depends on the final surface preparation of the cavities before the rf test \cite{KhanalIEEE}. The poor flux expulsion will result in higher flux sensitivity if the final surface preparation is other than electropolishing. When the RF surface is modified with impurity doping, the flux trapping sensitivity increases \cite{dhakal20}. Thus, efficient flux expulsion is required to minimize the additional rf loss related to the trapped flux. It is possible to achieve the full flux expulsion limit on cavities fabricated with cold work Nb after 800 $^\circ$C/3h heat treatment if one can maintain the temperature gradient between the cavities irises > 0.25 K/cm, which is not the case with cavities made from SRF grade Nb. However, the cavity treated at 800 $^\circ$C/3h requires larger temperature gradient (dT/ds > 0.2 K/cm) to achieve the flux expulsion ratio of $B_{sc}/B_n$ > 1.55, whereas with an additional 900 $^\circ$C/3h and 1000 $^\circ$C/3h heat treatment the same expulsion ratio can be achieve for dT/ds > 0.5 K/cm. The reason that larger temperature gradient required to achieve the maximum flux expulsion in SRF cavities showed that the recrystallization process could be different on deformed structures vs. the sample coupons. Additionally, significant number of magnetic flux pinning centers that are present in the micro-structure pinned the flux weakly and need higher thermal gradient to expel them away from the pinning centers. In fact, recent recrystallization experiments on 30 \% cold rolled Nb sheets showed a strain path dependence on recrystallization \cite{zach}.

To achieve the high flux expulsion ratio and hence the high quality factor in nitrogen doped SRF cavities, the cavities used in LCLS-II project were heat treated as high as 975 $^\circ$C \cite{ari} for some Nb sheet from Ningxia. The initial grain size specified with ASTM $\sim$ 5 showed better flux expulsion when treated at 900 $^\circ$C/3h compared to ASTM $\sim$ 7 requiring  975 $^\circ$C/3h to achieve the same level of flux expulsion. The electron back scattered imaging on sample coupons showed that the full recrystallization and grain growth occur after 800 $^\circ$C/3h with average grain size of 141$\pm6 \mu$m. However, some smaller grains of the order of 30 $\mu$m were observed along the cross-section of the sample within 3 mm sheet thickness. The sample subjected 900 $^\circ$C/3h has average grain size of 211$\pm13~\mu$m with an absence of smaller grains after the 900 $^\circ$C/3h \cite{shreyassrf23}. It was previously demonstrated that flux exit is harder in the regions of fine grins ($< 50 ~\mu$m) than the regions of larger grains ($\ge$ 100 $\mu$m) \cite{shreyas}. 

The higher flux trapping sensitivity in TE1-06 cavity compared to TE1-05 after initial 800 $^\circ$C/3h, even with better flux expulsion ratio suggest the presence of higher pinning center within the RF penetration depth. This results is in consistent with the presence of smaller grains size on sample coupons after 800 $^\circ$C/3h \cite{shreyassrf23}. Furthermore, it confirms that the flux expulsion depends on the bulk micro-structure, whereas flux trapping sensitivity depends on the details of surface preparations within the RF penetration depth.

\section{Summary}
There is a correlation of flux expulsion ratio and flux trapping sensitivity with the grain size, recrystallization as a result of high temperature heat treatment. The recrystallization study on the cut out of cavity half cell to understand the recrystallization with respect to the strain applied during cavity fabrications. New cavities are being fabricated with know percentage of cold work to further understand the controlled recrystalization to the flux expulsion and flux trapping sensitivity.

\section{ACKNOWLEDGEMENTS}
 We would like to acknowledge Dr. Gianluigi Ciovati and Dr. Shreyas Balachandran for several discussion. We would like to acknowledge Jefferson Lab technical staff members for the cavity fabrication, processing, and cryogenic support during RF test.

%
%
\ifboolexpr{bool{jacowbiblatex}}%
	{\printbibliography}%

\begin{thebibliography}{99} 
\bibitem{hasan}
H. Padamsee, " 50 years of success for SRF accelerators-a review", \textit{Supercond. Sci. Technol.}, \textbf{30}, 053003 (2017).
\bibitem{gigialex08}
G. Ciovati, and A. Gurevich, "Evidence of high-field radio-frequency hot spots due to trapped vortices in niobium cavities", \textit{Phys. Rev. ST Accel. Beams} \textbf{11}, 122001 (2008).
\bibitem{khanalsrf23}
B. D. Khanal, P. Dhakal, and G. Ciovati, "Quench Detection in a Superconducting Radio Frequency Cavity with Combined Temperature and Magnetic Field Mapping", this workshop, paper: SUSPB016.
\bibitem{dhakal20}
P. Dhakal, G. Ciovati, and A. Gurevich, "Flux expulsion in niobium superconducting radiofrequency cavities of different purity and essential contributions to the flux sensitivity", \textit{Phys. Rev. ST Accel. Beams}, \textbf{23}, 023102 (2020). 
\bibitem{sam}
S. Posen \textit{et al.,} "Role of magnetic flux expulsion to reach $Q_0 >3 \times 10^{10}$in superconducting rf cryomodule", \textit{Phys. Rev. Accel.Beams}, \textbf{22}, 032001 (2019).
\bibitem{dan1}
D. Gonnella, J. Kaufman, and M. Liepe, "Impact of nitrogen doping of niobium superconducting cavities on the sensitivity of surface resistance to trapped magnetic flux", \textit{J.Appl. Phys.} \textbf{119}, 073904 (2016). 
\bibitem{martina1}
M. Martinello \textit{et al.,} "Effect of interstitial impurities on the field dependent microwave surface resistance of niobium", \textit{Appl.Phys. Lett.} \textbf{109}, 062601 (2016). 
\bibitem{sam1}
S. Posen \textit{et al.,} "Efficient expulsion of magnetic flux in superconducting radiofrequency cavities for high $Q_0$ applications", \textit{J. Appl. Phys.} \textbf{119}, 213903 (2016). 
\bibitem{shreyas}
S. Balachndran \textit{et al.,} "Direct evidence of microstructure dependence of magnetic flux trapping in niobium", \textit{Sci. Rep.} \textbf{11}, 5364 (2021). 
\bibitem{antoine}
C. Z. Antoine, "Influence of crystalline structure on rf dissipation
in superconducting niobium" \textit{Phys. Rev. Accel. Beams} \textbf{22}, 034801 (2019).
\bibitem{gigi14}
G. Ciovati, P. Dhakal, and A. Gurevich, "Decrease of the surface resistance in superconducting niobium resonator cavities by the microwave field", \textit{Appl. Phys. Lett.} \textbf{104} (9), 092601 (2014). 
\bibitem{Khanalnapac22}
B. D. Khanal, S. Balachandran, P. J. Lee, S  Chetri, and P. Dhakal,"Magnetic flux expulsion in superconducting radio-frequency niobium cavities made from cold worked niobium",
\textit{ in the Proc. of NAPAC22}, Albaquarque, New MExico, USA. 
\bibitem{KhanalIEEE}
B. D. Khanal and P. Dhakal, "Insight to the Duration of 120 $^{\circ }$C Baking on the Performance of SRF Niobium Cavities," \textit{IEEE Transactions on Applied Superconductivity} \textbf{33} 3, 1-6 (2023). 
\bibitem{zach}
Z. L. Thune, C. McKinney, N. G. Fleming, and T. R. Bieler, "The influence of strain path and heat treatment variations on recrystallization in cold-rolled high-purity niobium polycrystals" \textit{IEEE Transactions on Applied Superconductivity}, \textbf{33}(5), 1-4 (2023). 
\bibitem{ari}
 A. D. Palczewski, D. Gonnella, O. S. Melnychuk, and D. A. Sergatskov, “Study of Flux Trapping Variability between Batches of Tokyo Denkai Niobium used for the LCLS-II Project and Subsequent 9-cell RF Loss Distribution between the Batches”, \textit{in Proc. SRF'19}, Dresden, Germany, Jun.-Jul. 2019, pp. 570-575. 
\bibitem{shreyassrf23}
S. Balachandran \textit{et al.,} "Microstructure development in a cold worked SRF niobium sheet after heat treatments", this workshop, paper ID: MOPMB041.










	\end{thebibliography}
	{%
	

} 
%
%


\end{document}